    \documentstyle[nato,psfig]{crckapb}

\def\ms{M$_{\odot}$}

\begin{opening}
\title{PLANETARY NEBULAE AS PROBES 
       OF STELLAR EVOLUTION AND POPULATIONS}

\author{LETIZIA STANGHELLINI}
\institute{Space Telescope Science Institute\\
          Affiliated to the Astrophysics Division,\\
Space Science Department of ESA\\
e-mail: lstanghe@stsci.edu}

\end{opening}

\runningtitle{THE CRCKAPB STYLE FILE}

\begin{document}


\begin{abstract}
Planetary Nebulae (PNe) have been used satisfactory to test
the effects of stellar evolution on the Galactic chemical environment.
Moreover, a link exists between nebular morphology 
and stellar populations and evolution. 
We present the latest results on Galactic PN 
morphology, and
an extension to a distance unbiased 
and homogeneous sample of Large Magellanic Cloud PNe.
We show that
PNe and their morphology may be successfully used as probes of
stellar evolution and populations.

\end{abstract}

\section{Relevance of Planetary Nebulae}

Planetary Nebulae represent the fate of most stars of masses 
in the range 1--10 \ms. The effects and consequences of their 
evolution are of great importance in stellar population studies, since
PNe represent the progeny of the most common stars in galaxies. 

PNe have been observed for centuries in the Galaxy, and more recently
in other galaxies such as the Magellanic Clouds, and between galaxies.
They have been identified as far as in the Virgo group. 
Planetary Nebulae studied in different galaxies are of exceptional interest
to probe stellar populations in a variety of environments. 
These studies would also allow the direct 
comparison of similar stellar populations in the different host galaxies
(see also Maciel, this volume).

\begin{figure}
\vspace{0.2cm}  
 \psfig{figure=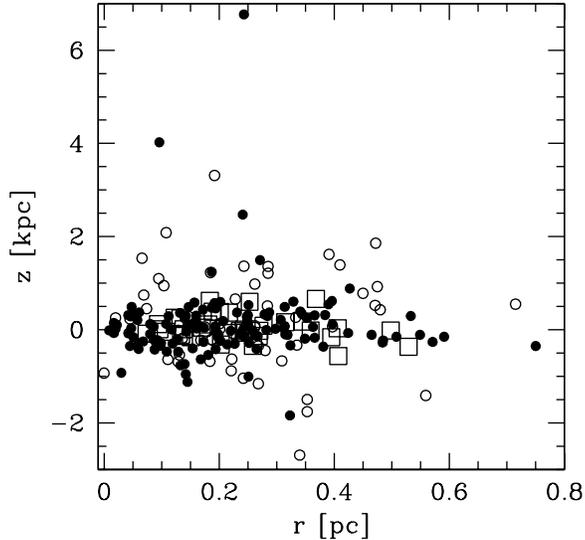,width=4in,height=4in}
\caption{Distribution of sizes and location within the Galaxy of round (open circles),
elliptical (filled circles), and bipolar (squares) PNe.}
\end{figure}

The ejection of the stellar envelopes of Asymptotic Giant Branch (AGB) 
stars yields to the PN formation, and also marks the beginning of the 
{\it cosmical recycling} process. In fact, PNe are among the
most important cosmic sources of carbon and other elements whose 
abundance is modified in the interiors of low- and intermediate-mass
stars. The carbon, oxygen, nitrogen, and helium abundances of PNe are
excellent tracers of the evolutionary paths of their progenitors, and their
initial mass (Renzini \& Voli 1981, Iben \& Renzini 1983). 
It is worth mentioning that the chemical
yields for AGB stars are at variation with respect to the initial chemical environment (van den Hoek \& Groenewegen 1997), and that rotation and other stellar effects may severely affect the
{\it classical} view of elemental production in AGB stars (e.g., 
Charbonnel 1995).

Planetary Nebulae also eject into the interstellar medium material
that has not been processed during the life of the low-mass star, such
as argon, neon, and sulphur. These
elements are uncorrupted by the evolution of low- and intermediate-mass
stars, thus their
abundance mix allows to infer the chemical conditions
of the interstellar medium (ISM) at the time of the formation of the
PN progenitor. 

Planetary Nebulae appear in a variety of morphologies, which have been
related to their chemical content (Peimbert 1978, Maciel 1989, 
Torres-Peimbert \& Peimbert 1997), and to stellar populations in the Galaxy (e.g., Stanghellini et al. 1993). It is thus important
to study the morphology of PNe for an independent test on stellar
evolution and populations.

Most of relations between PN morphology versus the evolution of the 
progenitor stars, and stellar populations, have been obtained for
Galactic PNe. Only the use of the HST cameras allows the spatial resolution
of extragalactic PNe. 
In $\S$2 we discuss the work on Galactic PN morphology, based
on contemporary databases. These results may be biased by the 
large uncertainties of the Galactic PN distance scale (Terzian 1993), 
and with the selection against PNe of low surface brightnesses, due to 
Galactic dust absorption.  
To circumvent these difficulties, and to study 
PNe in a different chemical environment, we have embarked a 
Large Magellanic Cloud PN survey, 
of which we give a preview in $\S$3. 
Section 4 includes a summary
and drafts future developments.

\section{Morphology of Galactic Planetary Nebulae}

Galactic PN morphology has been related to nebular and 
stellar physical properties in the past few decades (a review in Stanghellini 1995).
The importance of these investigations is two-fold. First, the correlations
between PN morphology and central star evolution allow us to use morphology to
infer the evolutionary paths of the stellar progenitors (Stanghellini et al. 1993;
Stanghellini \& Pasquali 1995). Second, by relating the PN morphology to kinematic,
spatial, and chemical quantities it is possible to use morphology as tracer of stellar 
populations (Peimbert 1978, Stanghellini et al. 1993). 

\begin{figure}
\vspace{0.2cm}  
 \psfig{figure=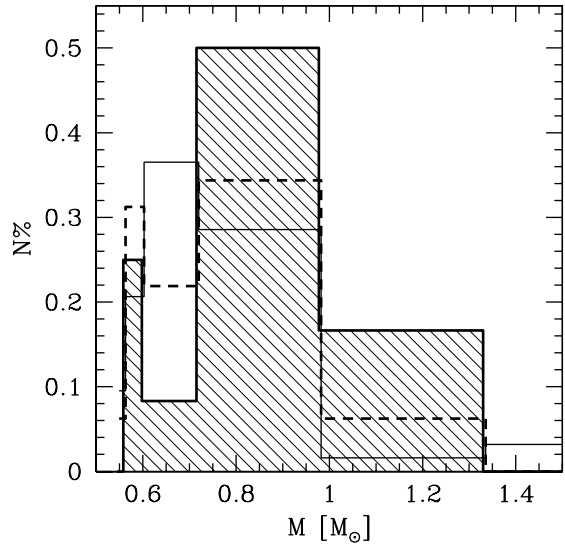,width=4in,height=4in}
\caption{Mass histogram for the central stars of round (thin line), elliptical (dashed
line), and bipolar (thick line and shaded histogram)
planetary nebulae.}
\end{figure}

In Figure 1
we show the distribution of Galactic PN as a function of their physical radii and the
altitude above (or below) the Galactic plane. The various morphologies are 
coded with different symbols. It is evident that most bipolar PNe
(sample from Manchado et al. 1996) lie closer to the Galactic plane than the
round and elliptical PNe, leading to the conclusion that bipolar PNe are the progeny of a young stellar population. This result is not new;
since the seventies Greig (1972), 
 found that {\it binebulous} PNe constitute
a younger PN population (Peimbert's {\it type I} PNe). The importance of this plot is that 
it is based on
a homogeneous and complete PN sample, placing this conclusion on firmer ground.

In Figure 2 we examine the mass distribution of central stars of PNe
across morphology. This plot is based on the same sample as Figure 1,
and on new  Zanstra analysis (Stanghellini et al. in preparation),
and it confirms what already found by Stanghellini et al. (1993):
bipolar PNe host stars in a wide range of masses, while very few high-mass central
stars are found within the round PNe. One again, this is indicative of two different 
stellar populations, and it is not in disagreement with the findings by
Peimbert and his collaborators.

\section{Morphology of Magellanic Cloud Planetary Nebulae}

The importance of the results discussed in the previous session is somewhat dimmed
by the difficulty to establish a Galactic distance scale to PNe. Even in the
most optimistic view, the distance to Galactic PNe is known to within 50$\%$, 
thus only the statistical distances to the selected samples
have a scientific significance, while the individual distances are very strongly biased.

To alleviate this problem we are observing a large sample of 
Large Magellanic Cloud (LMC) PNe (Cycle 8; Program ID: 8271; PI: Stanghellini; Co-Is: Shaw, Blades, Balick).
The main goal of our program is to detect LMC PN morphology and find the
connection to the physical conditions of the 
nebulae and their central stars. We
obtain the STIS/HST data in slitless mode. The 
advantage of slitless spectroscopy can be summarized in the following points: (1) each
exposure gives spatial and spectral information at once, obtaining {\it narrow band}
PN images plus excitation/ionization information and velocity information all at the same 
time; (2) we also contemporary obtain the spectrum of the central star, if bright enough to be visible.
The caveat of slitless spectroscopy is that spatial and spectral information may overlap 
for PNe larger than the spectral segregation between nearby lines. For example, this
technique can not be used for PNe larger than 1.5 arcsec in diameter.
Interestingly enough, these observations are almost ideally suited for LMC PNe, which typically span 0.8 arcsec in size.

From the space-based observations we can thus derive 
(a) the PN shape and size in 
the prominent emission lines (H$\alpha$, H$\beta$, [O III] $\lambda$5007, [N II], etc);
(b) the stellar and nebular spectrum, stellar continuum, central star identification, 
luminosity and temperature. 
Our HST program is about half
completed, the other half to be finished after the third HST servicing mission (SM3a).
Among the LMC PNe observed so far, we were able to recognize all the morphological classes that we were familiar with in Galactic PN studies.
 
From a related ground-based program (with NTT/EMMI at ESO)
we will derive detailed abundance and plasma diagnostics. At the time of writing, the ground-based observations (acquired in October 1999)
are completed but not analyzed yet, thus we use the literature abundances 
(main source: Leisy \& Dennefeld 1996) for the preliminary analysis shown 
in this paper. 

By correlating the PN morphology and chemical abundance we  find that the yields of the products of
stellar evolution change across morphology. We know that
the carbon, nitrogen, and oxygen abundance of PNe vary with progenitor stellar mass. In particular, carbon is produced during core
helium burning, but it is partially converted into nitrogen only in the massive
progenitors (e.g. van den Hoek \& Groenewegen 1997). The oxygen yield also 
varies with mass and other evolutionary parameters, but it seem to be consistently stable at the LMC metallicities 
(see Fig. 1 in van den Hoek \& Groenewegen 1997). We thus can use oxygen as a 
reference element for the discussion on the evolutionary enrichment.

\begin{figure}
\vspace{0.2cm}  
 \psfig{figure=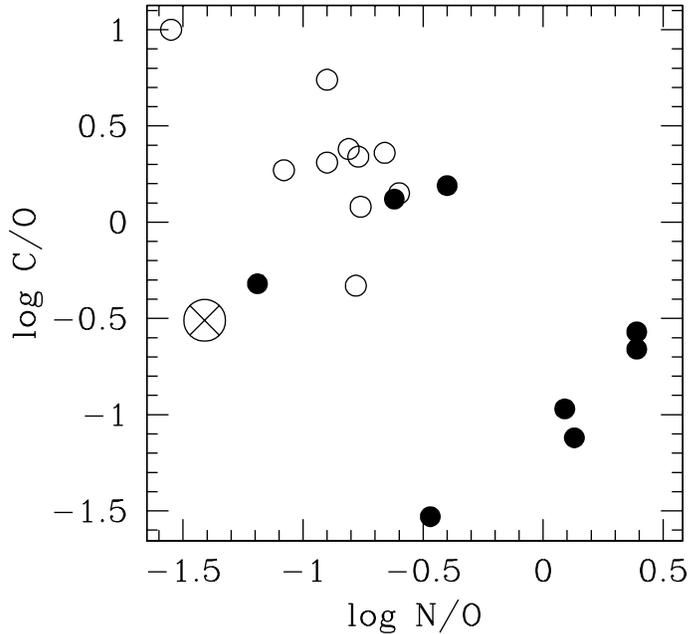,width=4in,height=4in}
\caption{Log C/O versus log N/O in LMC PNe. Filled symbols: asymmetric PNe (bipolar and quadrupolar); open symbols: symmetric PNe (round and 
elliptical). The crossed larger circle indicates the LMC average
of the H II regions.}
\end{figure}

Figure 3 shows the evolution of the C/O versus N/O in 
LMC PNe across morphological types. We see that almost all targets are 
enriched in N/O with respect to the local environment.
It is also evident that most symmetric PNe have high C/O and low N/O
ratios, and asymmetric PNe behaving the opposite way. This figure
give us a lot of confidence in what already found in Galactic PNe:
asymmetric PNe may have more massive progenitors that symmetric PNe.

 \begin{figure}
\vspace{0.2cm}  
 \psfig{figure=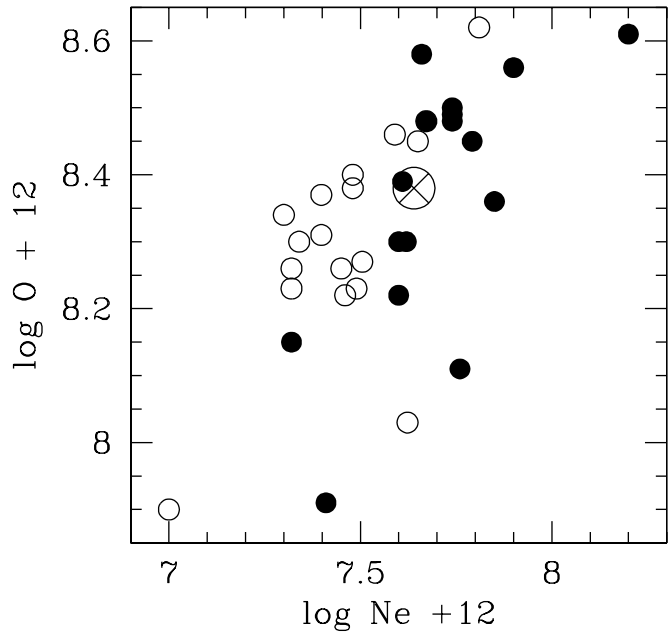,width=4in,height=4in}
\caption{ 
Oxygen versus Neon abundances in LMC PNe. Symbols as in Figure 3.}
\end{figure}

Figure 4 shows the marked difference of population-tracking elements
between symmetric and asymmetric PNe. Since both oxygen and neon should not change during evolution through the PN phase, both elements
describe the chemical environment prior to the formation of
the PN progenitors. Even with relatively small number statistics, we 
can infer that asymmetric PN progenitors have formed in an environment
that is chemically enriched 
with respect to where the progenitor of symmetric PN
originate. 
Most asymmetric PNe have a higher neon 
content than the local H II regions, while only one symmetric PN
is markedly enriched in neon with respect to the H II regions. 
Most of the symmetric PNe have lower oxygen abundance than
their surroundings, while more than half asymmetric PNe are
oxygen rich, again with respect to the local average.
Since there is such a sharp contrast in the abundance of
{\it primordial} elements across
morphological classes, we also test whether there are spatial segregations 
of either morphological class or chemistry in the LMC.
We did not find any convincing segregation. Our result
indicates that the (low- and intermediate-mass) 
stellar populations are mixed within the LMC.

\section{Summary}

By analyzing Galactic and LMC PNe, 
we searched for the relations between morphological class and 
chemical and spatial characteristics to test whether PNe can be used as
probes of stellar evolution and populations.
The previously known correlations of Galactic PN show a marked segregation of 
high-mass stars to form asymmetric PNe,
which is in agreement with the distribution of these PNe within the
Galactic disk. 
Novel studies in the Large Magellanic Clouds hint to connection
between morphology and stellar evolution, and formation/population
as well. The extragalactic studies are based on a database in the making.
The preliminary results lead to the
existence of two PN populations, distinct in their masses and chemical contents,
that can be resolved though morphological analysis. 

Ours future plans include the completion of our Cycle 8 HST program, 
and of the ground-based spectroscopy, to constitute a 
new, homogeneous, chemical/morphological database. 
We will derive the morphological
sequences and the dynamical ages, the stellar masses and 
evolutionary ages, the abundances, and finally we will model the surface brightness decline for each morphological class, as described in
Stanghellini et al. (1999).

We also hope to embark a similar study in the Small Magellanic Cloud (SMC),
in order to extend the
{\it chemical baseline} for PN formation to a low-metallicity environment.

\section{Acknowledgements}

I thank Dick Shaw, Arturo Manchado, and Eva Villaver for contributing substantially
to the science included in this paper, and Max Mutchler for the HST data analysis.
I whish to express my gratitude to Francesca Matteucci for 
inviting me to an outstanding meeting.

\end{document}